\documentclass[12pt,preprint]{aastex}
\usepackage{}
\usepackage{longtable}
\usepackage{graphicx}
\usepackage{subfigure}
\usepackage{longtable}
\usepackage{rotating}
\usepackage{lscape}
\usepackage{multirow}
\usepackage{array}
\usepackage{threeparttable}

\def\kepler{\textit{Kepler}}

\begin{document}

\title{ASTEROSEISMIC INVESTIGATIONS OF THE BINARY SYSTEM HD 176465}

\author{Ning Gai\altaffilmark{1,2}{$\star$},
Sarbani Basu\altaffilmark{2},
Yanke Tang\altaffilmark{1,2}}

\altaffiltext{1}{College of Physics and Electronic information, Dezhou University, Dezhou 253023, China}
\email{ning$_{-}$gai@163.com}

\altaffiltext{2}{Astronomy Department, Yale University, P.O.Box 208101, New Haven, CT 065208101, USA}


\altaffiltext{$\star$}{Corresponding author, email: ning$_{-}$gai@163.com; sarbani.basu@yale.edu}

\begin{abstract}
HD 176465 is a binary system whose both components are solar-like
pulsators and  whose oscillation frequencies were
 observed by the \kepler\ mission. In this paper we have modeled
the asteroseismic and spectroscopic data of the stars, and have determined
their convection-zone helium abundances using the signatures left by
the He{\sc ii} ionization zone on the mode frequencies. As expected
we find that the components of the binary have the same age within
uncertainties ($3.087 \pm 0.580$ Gyr and $3.569 \pm 0.912$ Gyr);
they also have the same initial helium abundance (Y$_{\mathrm{init}}$=0.253 $\pm$ 0.006 and
0.254 $\pm$ 0.008). Their current
metallicity ([Fe/H]=$-0.275 \pm 0.04$ and $-0.285 \pm 0.04$) is also the
same within errors.
Fits to the signature of the He{\sc ii}
acoustic glitch yields current helium abundances
of $Y_{\rm A} = 0.224 \pm 0.006$ and $Y_{\rm B} = 0.233 \pm 0.008$ for the
two components. Analyzing the complete ensemble of models generated
for this investigation we find that both the amplitude
and acoustic depth of the glitch signature arising from the second helium
ionization zone and the base of the convection zone (CZ) are functions of
 mass. We show that the acoustic depths of these glitches are
positively correlated with each other. The analysis can help us to detect the
internal structure and constrain the chemical compositions.
\end{abstract}


\section{Introduction}
The asteroseismic properties of the binary system HD~176465 were
observed by \kepler. It is one of the
few binary systems with detected solar-like oscillations in both
components. Binaries are important for studying stellar structure
and evolution because they are coeval and have the same
initial chemical composition. These systems are also an important means to
study the robustness of asteroseismic modelling. In this paper we use
the \kepler\ data as analyzed by White et al. (2017) to model the two
components of the binary, and we analyze the acoustic glitch signature
of the stars to determine their current convection-zone (CZ) helium abundances.

Acoustic glitches are regions where the structure (mainly sound-speed) of
a star changes over a length-scale shorter than the wavelength of the
modes. In stars, there are two major acoustic glitches,
the interface between convection and radiative zones, and the
second ionization zone of helium. The
change of the temperature gradient at the  base of the convective envelope, and the
depression in the first adiabatic index $\Gamma_{1}$ in the second helium
ionization zone cause localized variations of the sound speed, which in turn
affect the frequencies of acoustic modes in these regions. These localized
features introduce an oscillatory component in the
frequencies as a function of radial order $n$ (Gough \& Thompson 1988;
Vorontsov 1988; Gough 1990). This oscillatory component is proportional to

\begin{equation}
\sin(2\tau_{g}\omega_{n,l}+\phi), \hspace{1.0cm} \tau_{g}=\int^{R}_{r_{g}}\frac{dr}{c},
\label{eq:glitch}
\end{equation}
where $\omega_{n,l}$ is the angular frequency of a mode with degree $l$ and
radial order $n$, and $\phi$ is a phase. $\tau_{g}$ is the acoustic depth of
the glitch, which is calculated from the location of glitch to the surface of
the star of radius $R$. The quantity  $c$ is the sound speed, and
$r_{g}$ is the radial distance of the glitch.

In the earliest studies, the oscillatory signature has been used to
determine the extent of overshoot below convection zone of the Sun
(see e.g., Gough 1990;
Basu et al. 1994; Monteiro et al. 1994; Roxburgh \& Vorontsov 1994;
Christensen-Dalsgaard et al. 2011). Additionally, the acoustic glitch has
been studied theoretically to determine if they could be used to estimate the
location of the base of the convective envelope and the second helium
ionization zone (Monteiro et al.  2000; Mazumdar \& Antia 2001;
Roxburgh \& Vorontsov 2003; Houdek 2004; Houdek \& Gough 2006, 2007, etc.)
in other stars using only low-degree modes.
Recently, with the high-precision seismic frequencies
observed by \kepler\ and CoRoT space missions, Miglio et al. (2010)
studied a red giant to determine the position of He{\sc ii} ionization zone using
the modulation of the frequency separations observed by CoRoT. Mazumdar et
al. (2012) also use the CoRoT data to detect the acoustic depth of the base
of the convection zone and the second helium ionization zone for a main sequence star
HD~49933. Then using the oscillation data from \kepler, Mazumdar et al. (2014)
measured the acoustic depths of both glitches for 19 stars. Verma et al.
(2017) analyzed  both glitches for 66 stars in the \kepler\
seismic LEGACY sample (Lund et al. 2017).
An accurate determination of the position of the glitch at the convection-zone
base is helpful for understanding stellar dynamo processes in cool stars as well as
details of stellar structure.

In addition to the study of the position of the glitch,  the amplitude
of the glitch signature caused by the depression in $\Gamma_{1}$  in the He{\sc ii}
ionization zone
can be used to determine the helium abundance in the stellar
envelopes of  low mass cool stars. Helioseismic data have been
used successfully to determine the helium abundance in the solar
envelope (see e.g., Basu \& Antia 1995, etc.), and in a theoretical study
Basu et al. (2004) showed that the signature
of the He{\sc ii} acoustic glitch in low-degree mode frequencies of
 main sequence stars could be used to determine helium in the envelopes
of these stars.  This technique
was then used by  Verma et al. (2014a), who, using
\kepler\ observations,
determined the helium abundance of both components of the solar-analog binary
system 16 Cyg.

In this work we first determine the mass, radius and age of the two components
of the HD 176465 system, and also
we estimate the helium abundance of both components  using oscillation
frequencies observed by the \kepler\ satellite. We then
analyze the glitch signatures from
the base of the convective envelope and the second helium ionization zone
to determine the acoustic depth of the base of the convection zone and
the helium ionization zone. The rest
of the paper is organized as follows:
we describe the observed parameter of HD 176465, which
include the individual frequencies and spectroscopic parameters used for
asteroseismic modelling in \S~\ref{sec:modeling}. In this section, we also
present the procedure to calibrate the best-fit model for both components. We
describe the technique that used to estimate the helium abundance and the
position of the second helium ionization zone and the base of the convective
envelope from the signature of the acoustic glitch in \S~\ref{sec:glitches}.
Finally, we discuss our conclusions in \S~\ref{sec:conc}.

\section{Data and Analysis}
\label{sec:modeling}

\subsection{Observational Constraints on HD 176465}
\label{sec:observe}

To determine the stellar properties and conduct the asteroseismic analysis of acoustic
glitch in the interior, we adopt the set of individual frequencies and
spectroscopic parameters published by White et al. (2017). HD 176465 was observed
in short-cadence mode (SC; 58.85s sampling) for 30 days with high-quality photometric
observations by the \kepler\  during the asteroseismic survey phase (from 20-07-2009 to
19-08-2009, i.e., Quarter 2.2), and continuously from the end of the survey (37 months,
Quarters 5 - 17). Additionally, the system was observed in the long-cadence mode (LC;
29.43 min sampling) for the entire nominal mission (4 years, Q0 - Q17), although this
sampling is not rapid enough to sample frequencies of the solar-like oscillations,
which are well above the LC Nyquist frequency. Jenkins et al. (2010) and Garc\'{\i}a et al. (2011)
were described how the SC time series were prepared from raw observations and corrected
to remove outliers and jumps respectively. The observed frequencies
include 16 consecutive orders for the $l = 0$ (radial), $l = 1$ (dipole) and $l = 2$
(quadrupole) modes for HD 176465~A, as well as 11 consecutive orders for the $l
= 0, 1, 2$ modes for HD 176465~B. The individual frequencies and their uncertainties
were shown in Table 4 and 5 of white et al. (2017). The large frequency separation are $146.79
\pm 0.12$ and $155.42 \pm 0.13\; \mu$Hz for HD 176465 A and B respectively
(see Table 1 of White et al. 2017).

In addition to the observed frequencies, the spectroscopic parameters are also
presented in Table 1 of White et al. (2017). A spectrum of HD 176465 was obtained from the
ESPaDOnS spectrograph by the 3.6m Canada--France--Hawaii Telescope. Different
spectroscopic methods were used to analyze the spectrum and derived the stellar
parameters and metallicity by Bruntt et al. (2012) who determined a surface
metallicity of [Fe/H]$ = -0.30 \pm 0.06$ dex.
Because the components in binary were formed from the
same gas cloud at the same time, we expect the both components have the same
initial metallicity. The observed effective temperatures are $T_{\rm eff} = 5830
\pm 90 $K for HD 176465 A and  $T_{\rm eff} = 5740 \pm 90 $K for HD 176465 B.
Besides the individual frequencies, the above metallicity and the effective
temperatures are used to constrain the models of HD 176465.

\subsection{Stellar Models}
\label{sec:model}
\subsubsection{Input Physics and Initial Input Parameters}
\label{sec:inputphy}
For this work, stellar models are calculated from the zero-age main sequence
(ZAMS) to the end of main sequence with the Yale Rotation and Evolution Code
(YREC; Demarque et al. 2008) in its non-rotating configuration. The input
physics includes the OPAL equation of state tables of Rogers \& Nayfonov
(2002), and OPAL high temperature opacities (Iglesias \& Rogers 1996)
supplemented with low temperature opacities from Ferguson et al. (2005). The
NACRE nuclear reaction rates (Angulo et al. 1999) are used. The gravitational
settling of helium and heavy elements use the formulation of Thoul et al.
(1994).  We use the Eddington $T$-$\tau$ relation. Convection is modeled using
the standard mixing length theory. An overshoot of $\alpha_{c}$ = 0.2$Hp$ is assumed
for models with convective cores.

For both components of HD~176465, we construct models for the same grid of
initial stellar parameters, but the stars are analyzed independently. These parameters cover masses in the
range M = 0.8 to 1.18 M$_{\odot}$ in steps of 0.02M$_{\odot}$. The values of
the mixing-length parameter $\alpha_{\mathrm{MLT}}$ = 1.6, 1.7, 1.826, 1.9, 2.0. Based on the
observation, we set the initial metallicities is $-0.20 \leqslant$ [Fe/H]
$\leqslant -0.40$ with a resolution of 0.05 dex. The metallicity [Fe/H] is
defined as [Fe/H] = log(Z/X)$_{star}$ -- log(Z/X)$_{\odot}$. In order to
convert the metallicity [Fe/H] to $Z$, we need the values of
solar metallicity (Z/X)$_{\odot}$. The solar metallicity mixture (Z/X)$_{\odot}$ = 0.023
(Grevesse \& Sauval 1998, hereafter GS98) had been used extensively for many years.
Although it is an old value, the standard solar models constructed with GS98 mixture
satisfied helioseismic constraints quite well (Basu \& Antia 2008). Recently, the
solar metallicity was revised to lower value, e.g. (Z/X)$_{\odot}$ = 0.018 (Asplund et al. 2009,
hereafter AGSS09), (Z/X)$_{\odot}$ = 0.0191 (Lodders 2009) and (Z/X)$_{\odot}$ = 0.0211
(Caffau et al. 2010). Compared to the GS98 mixture, the solar models constructed with
lower abundance disagree with the helioseismic inferred density and sound-speed profiles, etc.
(Serenelli \& Basu 2010; Basu 2016, Yang 2016). In order to estimate the effect of
different solar metallicities on the stellar models, we use both GS98
((Z/X)$_{\odot}$ = 0.023) and a lower metallicity AGSS09 ((Z/X)$_{\odot}$ = 0.018) mixture as a standard
for calibrating the metallicity of HD176465.

The initial helium for our evolutionary
models are determined using a helium-to-metal enrichment law $\Delta Y/\Delta Z$ = 1.4 on
the basis of the Big Bang nucleosynthesis primordial values Z$_{0}$ = 0.0 and Y$_{0}$ = 0.248
(Steigman 2010). Moreover, we also refer to the modelling results of initial helium for HD176465
by White et al.(2017) using different codes (see Table 2 and 3 in White et al. 2017). At last,
we set the initial value of helium ranging
from 0.250 to 0.266 and the step size is 0.004. Although both components of a binary
have the same age, we do not apply this constraint, but instead determine the
ages of the two components independently.

\subsubsection{The Surface Correction of Oscillations}
Based on the above input physics and different combinations of initial
parameters, we calculate a total of 5000 evolutionary tracks. For each model on
these tracks, we calculate the individual frequencies and compare them with the
observed frequencies. The low-$l$ frequencies of all models are calculated with the pulsation
code JIG (Guenther 1994). Because the near-surface layers of a star are unable to
be modeled properly, it will result in the differences of the frequencies
between the star and model, which is known ``surface term". It also make the
precise asteroseismic analyses to be difficult. In this work, we use the
two-term model of surface term to correct the frequencies of models. It was
proposed by Ball \& Gizon (2014) and works well in all parts of the HR diagram
(Schmitt \& Basu 2015). The correction of the frequencies is defined by
(Equations (4) of Ball \& Gizon 2014 and Equation (1) of Howe et al. 2017),
\begin{equation}
\mathcal{F}(\nu)=E_{nl}\delta\nu_{corr}=a_{cubic}(\frac{\nu}{\nu_{ac}})^{3}+a_{inv}(\frac{\nu}{\nu_{ac}})^{-1},
\label{eq:corr}
\end{equation}
where $a_{\mathrm{cubic}}$ and $a_{\mathrm{inv}}$ are fitting coefficients, and $\nu_{ac}$ = 5 mHz
is the acoustic cut-off frequency. $\mathcal{F}(\nu)$ is the scaled frequency differences,
which is fitted, as a function of frequency. It is easy to remove the dependence of
the frequency difference on inertia. In this work, we refer to the detailed description
of two-term linear least-squares fits which was given in Appendix A of Howe et al. (2017)
, and get the coefficients $a_{\mathrm{cubic}}$ and $a_{\mathrm{inv}}$ of models for HD 176465 that are
listed in Table~\ref{tab:bestfit}. For calculated frequencies, we use Equation(~\ref{eq:corr})
to calculate the corrections $\delta\nu_{corr}$, and get the corresponding corrected
frequencies of models $\nu_{corr}$ = $\nu_{mod}$ + $\delta\nu_{corr}$.

\subsubsection{$\chi^{2}$-minimization Procedure}
In order to get the best-fit models for HD 176465, we use
$\chi^{2}$-minimization method to compare the models with observations.
We define the reduced $\chi^{2}_{\nu}$ below to compare the observed
frequencies and the ones of models.
\begin{equation}
\chi^{2}_{\nu}=
\frac{1}{N}\sum_{i=1}^{N}\left(\frac{\nu_{\rm obs}-\nu_{\rm mod,corr}}{\sigma_{\rm obs}}\right)_{i}^{2},
\label{eq:chimu}
\end{equation}
where $\nu_{\rm mod,corr}$ are the surface-term corrected frequencies
of the model and $\sigma_{\rm obs}$ is the 1$\sigma$ uncertainty in
the observed frequencies. Additionally, we also calculate the reduced
$\chi^{2}$ for all  models using both asteroseismic and spectroscopic
constraints:
\begin{equation}
\chi^{2}=\frac{1}{N}\sum_{i=1}^{N}\left(\frac{q_{\rm obs}-q_{\rm mod}}{\sigma_{\rm obs}}\right)_{i}^{2},
\label{eq:chi}
\end{equation}
where $q_{\rm obs}$ represents the observed parameters ($T_{\rm eff}$,
[Fe/H], frequencies $\nu$), $q_{\rm mod}$ are the corresponding
parameters of stellar models and $\sigma_{\rm obs}$ denotes the
1$\sigma$ uncertainty in the observations.

\begin{figure}
\epsscale{1.0} \plotone{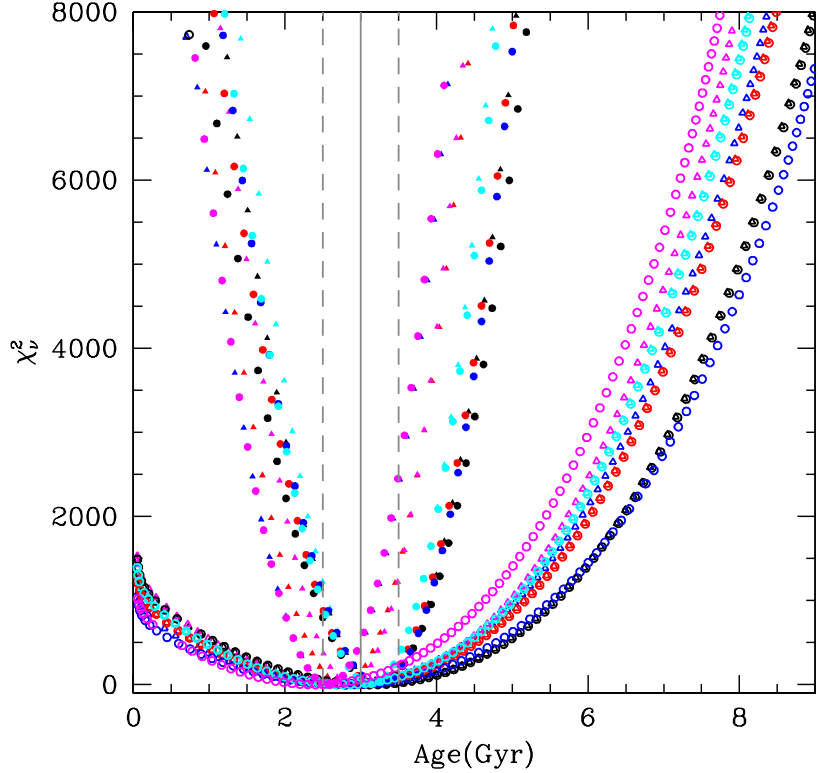} \caption{$\chi_{\nu}^{2}$ vs. age for
the evolutionary tracks with different mixing-length parameters
$\alpha_{\mathrm{MLT}}$ and different solar metallicity mixtures. Each
evolutionary track has one and only one $\chi_{\nu}^{2}$-minimization
model that has been listed in Table~\ref{tab:bestfit}. Different
shape symbols represent different solar mixture and different
colours represent different $\alpha_{\mathrm{MLT}}$. Circles and triangles
show the tracks with GS98 and AGSS09 respectively. The black, red,
blue, cyan and magenta colours represent tracks with $\alpha_{\mathrm{MLT}}$ =
1.6, 1.7, 1.826, 1.9, 2.0 respectively. The solid circles and triangles
show the tracks for HD176465 A, and the open circles and triangles
show the tracks for HD176465 B. The solid gray vertical line
indicates 3.0 Gyr, while the gray dashed lines indicate errors of 0.5
Gyr.}  \label{fig:chit}
\end{figure}

In order to get the best-fit model, non-asteroseismic parameters and
asteroseismic parameters (mainly individual frequencies) are usually
used together to obtain a total residual error $\chi^{2}$ (e.g.
Lebreton 2012, Lebreton \& Goupil 2014). Sometimes these
parameters are used separately. Firstly, the non-asteroseismic
parameters are used to narrow the range of models and then the
asteroseismic parameters are used to get the best models.
Owing to the high precision space-based observations, dozens of
oscillation frequencies can be get for solar-like stars.
Additionally, the uncertainty in the frequencies are usually
much smaller than those in non-asteroseismic parameters. Some
researches have been done to optimize the precision of asteroseismic
inferences by matching the ratios of frequency separations or
the individual frequencies (Metcalfe et al. 2014; Silva Aguirre
et al. 2013). These methods improve the precision by a factor
of two or more over the techniques that only use the global
oscillation parameters (e.g. Lebreton \& Goupil 2014; Silva
Aguirre et al. 2013; Mathur et al. 2012). Recently, Wu \& Li
(2016, 2017) only use the observed high-precision oscillation
frequencies to constrain theoretical models and determine the
fundamental parameters precisely for the Sun and the solar-like
oscillator KIC6225718. In this work, we calculate $\chi_{\nu}^{2}$
for all the models and then choose the $\chi_{\nu}^{2}$-minimization
model on each evolutionary track as the best model for the
corresponding track. At last, from all the $\chi_{\nu}^{2}$-minimization
models, we select the model with minimum value of $\chi_{\nu}^{2}$
as the best-fit models for HD176465 A and B. For some models with
very similar $\chi_{\nu}^{2}$, we will refer to the non-asteroseismic
parameter and choose model with smaller $\chi^{2}$ as the better
candidate of the star. In Table~\ref{tab:bestfit}, we list the
fundamental parameters of 10 better candidate models MA1$\sim$MA10
for HD176465 A and MB1$\sim$MB10 for HD176465 B respectively.
Each candidate model with minimum value of $\chi_{\nu}^{2}$ is
selected from a total of 500 evolutionary tracks with different
initial combinations (M, Y$_{\mathrm{init}}$, Z$_{\mathrm{init}}$)
and a specific mixing-length parameters $\alpha_{\mathrm{MLT}}$
and the solar metallicity mixture $(Z/X)_{\odot}$. The evolutionary
tracks of these models in Table~\ref{tab:bestfit} are shown in
Figure~\ref{fig:chit}. From Figure~\ref{fig:chit}, we can see that
each track display regular variation of $\chi_{\nu}^{2}$ and
converge to a regular shape with a unique minimum $\chi_{\nu}^{2}$.
It ensure that only one model is selected as the best model for a
specific evolutionary track. Additionally, Figure~\ref{fig:chit}
show that the age of all the $\chi_{\nu}^{2}$-minimization models
are almost in a range 3.0$\pm$0.5 Gyr that is the modelling result
by White et al. (2017), in agreement with the gyrochronological
value $\sim$ 3Gyr.

Table~\ref{tab:bestfit} show that the metallicities of models
calibrated with AGSS09 are almost lower than the observed
metallicity -0.3$\pm$0.06. According to the above analysis,
model M8A, with minimum $\chi_{\nu}^{2}$ = 5.298 among 10
candidates MA1$\sim$MA10 in Table~\ref{tab:bestfit},
is selected as the final best-fit model for HD176465 A.
Furthermore, T$_{eff}$ and [Fe/H] of M8A also match the
non-asteroseismic observation within 1$\sigma$ error with
$\chi^{2}_{T_{eff}}<$ 1 and $\chi^{2}_{[Fe/H]}<$ 1. As we
know that, the shared formation history of binary star system
result in the same age and initial chemical composition for
both components of HD176465. According to Table~\ref{tab:bestfit},
the model M8B with the combination of
($\alpha_{\mathrm{MLT}}$, $(\mathrm{Z/X})_{\odot}$) = (1.826, 0.023)
(noting as $\alpha$1.826GS98), as same as M8A, is the best-fit
model of HD176465 B. Although $\chi_{\nu}^{2}$ of M8B is slightly
larger than that of some other models (e.g. M1B, M2B, M4B, M6B, M7B),
non-asteroseismic parameters of M8B are consistent with the observations
($\chi^{2}_{T_{eff}}<$ 1 and $\chi^{2}_{[Fe/H]}<$ 1). Additionally,
M8A and M8B have the same initial chemical abundance $Y_{\mathrm{init}}$
and $Z_{\mathrm{init}}$, surface metallicity [Fe/H] and age. Hence, we
select the models with the combination $\alpha$1.826GS98 to deeply
analyze the fundamental parameters and the interior structures of
HD176465 A and B.

\subsubsection{Modelling Results and Comparisons With the Work of White et al.(2017)}
\begin{figure}
\epsscale{1.0} \plotone{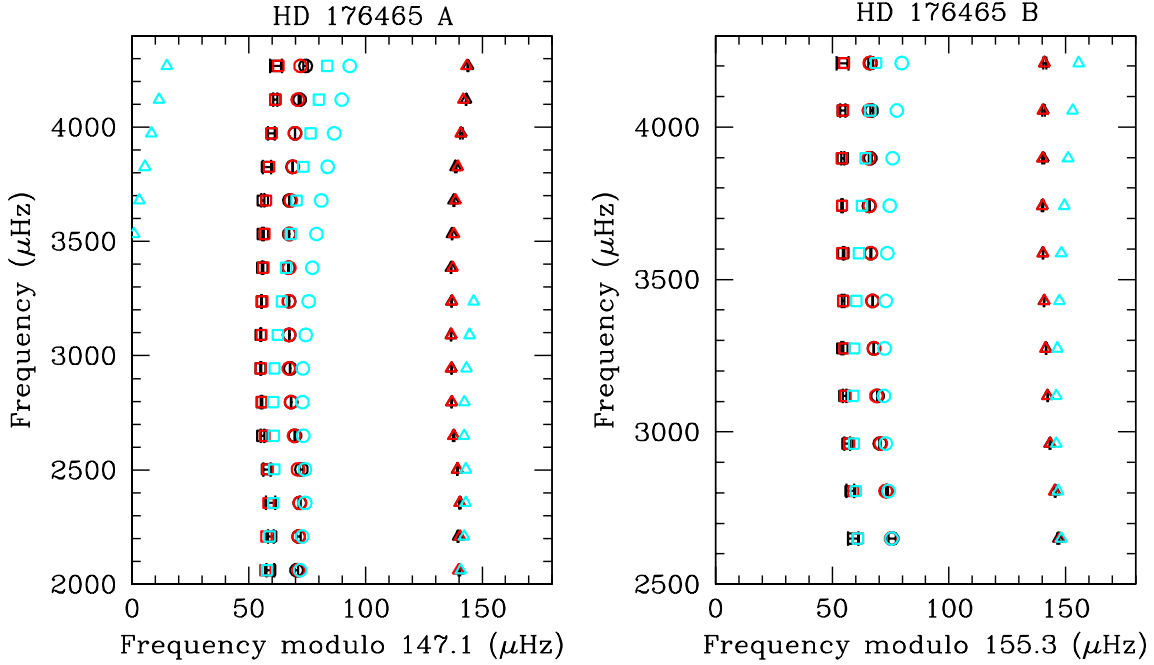}
\caption{\'{E}chelle diagrams for HD 176465 A (left) and HD 176465 B (right).
The surface corrected frequencies of best-fit models of M8A and M8B in
Table~\ref{tab:bestfit} are indicated by the red
symbols. The uncorrected frequencies of best-fit models are indicated by the
cyan symbols. The black points with error bars are the observed frequencies.
Different shapes of symbols represent the different modes: circles are $l$ = 0
modes, triangles are $l$ = 1 modes and squares are $l$ = 2 modes.}
\label{fig:elle}
\end{figure}

Figure~\ref{fig:elle} shows the echelle diagrams comparing the
observed and model frequencies and thereby illustrating the quality of
the match to the frequencies before and after surface-term correction using the
two-term model of Ball \& Gizon (2014). The large frequency separation are 147.1
$\mu$Hz and 155.3 $\mu$Hz for HD 176465 A and B respectively. They are
consistent with the observations within 1$\sigma$. The
large-frequency separations  were calculated with
surface-term  corrected frequencies.

Finally, we derive the stellar parameters as the likelihood
weighted mean and standard deviation from the models with minimum $\chi^{2}$ along
each stellar track (see Gilliland et al. 2011 for the technique).
\begin{equation}
<p>=\frac{\sum\limits_{i=1}^{n} \mathcal{L}p}{\sum\limits_{i=1}^{n} \mathcal{L}},
\label{eq:meanp}
\end{equation}
\begin{equation}
<\sigma_{p}>=(\frac{\sum\limits_{i=1}^{n} \mathcal{L}(p-<p>)^{2}}{\sum\limits_{i=1}^{n} \mathcal{L}})^{1/2},
\label{eq:erorrp}
\end{equation}
\begin{equation}
\mathcal{L}=(\prod\limits_{i=1}^{N}\frac{1}{\sqrt{2\pi}}\sigma_{i})\times\exp(-\chi^{2}/2),
\label{eq:likeli}
\end{equation}
where $p$ $\equiv$ [M, Y$_{\mathrm{init}}$, Z$_{\mathrm{init}}$, [Fe/H], age, R], $<p>$ is
the likelihood weighted mean value of the fundamental parameters, $<\sigma_{p}>$ is
standard deviation, $\mathcal{L}$ is the likelihood function and $\chi^{2}$ has been
defined by Equation(~\ref{eq:chi}). The modeling results for HD 176465 A and B
are presented in Table~\ref{tab:compareA} and ~\ref{tab:compareB}. We see that,
as expected, the two components of HD 176465 have the same initial composition.
From the models with combination $\alpha$1.826GS98,
we obtain that the current metallicities [Fe/H]$_{\rm A} = -0.275 \pm 0.04$ and
[Fe/H]$_{\rm B} = -0.285 \pm 0.04$ are quite similar and consistent with the
observation [Fe/H] = $-0.30 \pm$ 0.06 derived by Bruntt et al. (2012). The age of the two
stars $t_{\rm A} = 3.087\pm 0.580$Gyr and $t_{\rm B} = 3.569 \pm 0.912 $Gyr
are, within uncertainties, the same. The results also illustrate that
despite issues with surface uncertainties, asteroseismic analyses are reliable.

The detailed comparisons of fundamental parameters between our
results (YREC1, YREC2) and the results of White et al. (2017)
with different codes (AMP, ASTFIT, BASTA and MESA) are listed
in Table~\ref{tab:compareA} for HD176465 A and Table~\ref{tab:compareB}
for HD176465 B. According to Table~\ref{tab:compareA} and
~\ref{tab:compareB}, we can see that our results (YREC1) are
more similar with the results of ASTFIT and BASTA models. These
three modelling methods have almost the same input physics
(including the mixing length parameter, $\alpha_{\mathrm{MLT}}$)
and minimization technique. The YREC models matches individual
frequencies with surface effects correction to get the best-fit
model, which is the same as ASTFIT method, and the BASTA fits
frequency ratios. But the mass for the primary component obtained
with YREC1 is slightly larger than that of AMP, ASTFIT and BASTA
models, while the YREC1 model has a lower initial helium abundance.
This degeneracy between helium abundance and mass also shown in
the results obtained with MESA, which is well-known in asteroseismic
modelling of solar-like oscillations (Lebreton \& Goupil 2014;
Silva Aguirre et al. 2015). In Table~\ref{tab:compareA} and
~\ref{tab:compareB}, except ASTFIT adopted the Grevesse \& Noels
(1993, hereafter GN93) solar mixture, other codes used the GS98
solar mixture (Grevesse \& Sauval 1998). In this work, we also get
the likelihood weighted mean and standard deviation of fundamental
parameters (YREC2) from models with different $\alpha_{\mathrm{MLT}}$
and AGSS09 solar mixture. From Table~\ref{tab:compareA} and
~\ref{tab:compareB}, we can see that the mass of stars with AGSS09
is slightly lower than that of other codes. The initial metallicity
with AGSS09 is lowest in all the results. The current metallicities,
[Fe/H]$_{A}$ = -0.366 $\pm$ 0.018 and [Fe/H]$_{B}$ = -0.363 $\pm$ 0.025,
is poorer than the observations. For the results of White et al. (2017),
four different modelling codes converge on models with mean value of
mass M$_{A}$ = 0.95 $\pm$ 0.02 M$_{\odot}$, M$_{B}$ = 0.93 $\pm$ 0.02
M$_{\odot}$ and mean radius R$_{A}$ = 0.93 $\pm$ 0.01 R$_{\odot}$,
R$_{B}$ = 0.89 $\pm$ 0.01 R$_{\odot}$. The ages of both components
were found to be 3.0 $\pm$ 0.5 Gyr, which agree with the
gyrochronological value $\sim$ 3Gyr. Our results for the mass,
radius and age are consistent with those of White et al. (2017)
which were determined using different codes.

\section{Study of Acoustic Glitches}
\label{sec:glitches}

\subsection{The Technique}
\label{sec:technique}

The oscillatory signatures in the frequencies caused by
acoustic glitches are very small. The second differences of the frequencies
are usually adopted to amplify the glitch signal by suppressing
the smooth variation of the frequencies to a large extent, these are defined as:
\begin{equation}
\delta^{2}\nu_{n,l}=\nu_{n+1,l}-2\nu_{n,l}+\nu_{n-1,l}.
\label{eq:sed}
\end{equation}

Fitting the second differences is easier than the fit of frequencies
themselves to measure the oscillatory signal (Gough 1990; Basu et al. 1994,
2004; Mazumdar \& Antia 2001), however, neighbouring second-differences have
correlated errors, and hence the full error-covariance matrix must be taken into
account. The $\delta^{2}\nu$ are
fitted to a suitable functional form that represents the glitch signatures from
the base of the convective envelope and the second helium ionization zone. In
this work, we use one of the functional forms suggested by  Basu et al. (2004):
\begin{equation}
\delta^{2}\nu_{n,l}=a_{1}+a_{2}\nu+\frac{a_{3}}{\nu^{2}}\\
+\left(b_{1}+\frac{b_{2}}{\nu^{2}}\right)\sin(4\pi\nu\tau_{\rm He}+\phi_{\rm He})\\
+\left(c_{1}+\frac{c_{2}}{\nu^{2}}\right)\sin(4\pi\nu\tau_{\rm CZ}+\phi_{\rm CZ}),\\
\label{eq:fsed}
\end{equation}
where $a_{1}$, $a_{2}$, $a_{3}$, $b_{1}$, $b_{2}$, $\tau_{\rm He}$, $\phi_{\rm He}$,
$c_{1}$, $c_{2}$, $\tau_{\rm CZ}$, $\phi_{\rm CZ}$ are 11 free parameters. In
Equation(~\ref{eq:fsed}), the first three terms defines the smooth part of the
second differences, the fourth term represents the oscillatory signal from the
He{\sc ii} ionization zone, while the last term represents the signal from the base
of the convection zone. $\tau_{He}$ and $\tau_{CZ}$ are the acoustic depth of
the He{\sc ii} ionization zone and the base of the convection zone respectively.
$b_{1}+b_{2}/{\nu^{2}}$ and $c_{1}+c_{2}/{\nu^{2}}$ are the amplitudes of acoustic
glitch oscillation signature from the second helium ionization zone and the
base of the convection zone respectively. In Equation(~\ref{eq:fsed}), the 11
unknown parameters are determined by a nonlinear least-squares fit to the second
differences of the frequencies to get the parameters that give the lowest $\chi^2_{fit}$.
\begin{equation}
\chi^2_{fit}=\sum\limits_{i=1}^{N}[\frac{\delta^{2}\nu-\delta^{2}\nu_{fit}}{\sigma}]_{i}^{2},
\label{eq:chifit}
\end{equation}
where $\delta^{2}\nu$ is the un-fitted second differences of the observations or models,
$\delta^{2}\nu_{fit}$ is the fitted second differences, $\sigma$ is the uncertainty of
the $\delta^{2}\nu$ derived from individual frequencies by the error propagation.

In order to obtain the distribution of the fitted parameters, the first step of
this technique is generating 1000 Monte Carlo realizations of the frequencies
by adding different random perturbations to the actual frequencies with
specified standard deviation equal to 1$\sigma$ uncertainty in the actual
frequencies. The second step is getting 100 different initial guesses, which
are selected randomly in some range of possible values, to weaken the
dependence of the fitted parameters on the choice of initial guesses.
The different sets of initial guesses are obtained by randomly perturbing
a reasonable value for each free parameters. Because HD176465 are the solar-like
oscillators with the mass similar with the Sun, firstly, the typical values of
initial guesses are expected to be similar with the parameters for the Sun
(e.g.$\tau_{\rm He}\sim 700s$, $\tau_{\rm CZ}\sim 2300s$). Based on the first set
of initial guesses, another 99 sets of the initial parameters obtained by adding
random perturbations to the first set with the standard deviation equal to 20\% of
the first set of parameters. Then, for
each of 1000 realizations and actual frequencies, the fitting of
Equation(~\ref{eq:fsed}) are carried out for 100 times using different initial
guesses. At last, from the 100 trials, the fitting parameters with the minimum
value of $\chi_{fit}^{2}$ is accepted as the best fit value for that particular
realization.

For each of fitting parameters, we choose the median value from the
distribution of 1001 fitting values (for 1000 realizations and actual one) as
the value of the parameter. The upper 34\% and lower 34\% of the distribution on
either side of the median is the $\pm 1\sigma$ uncertainty of the parameter.

\subsection{Glitch Analysis of HD 176465}
\label{sec:hd176465}

\begin{figure}
\epsscale{1.0} \plotone{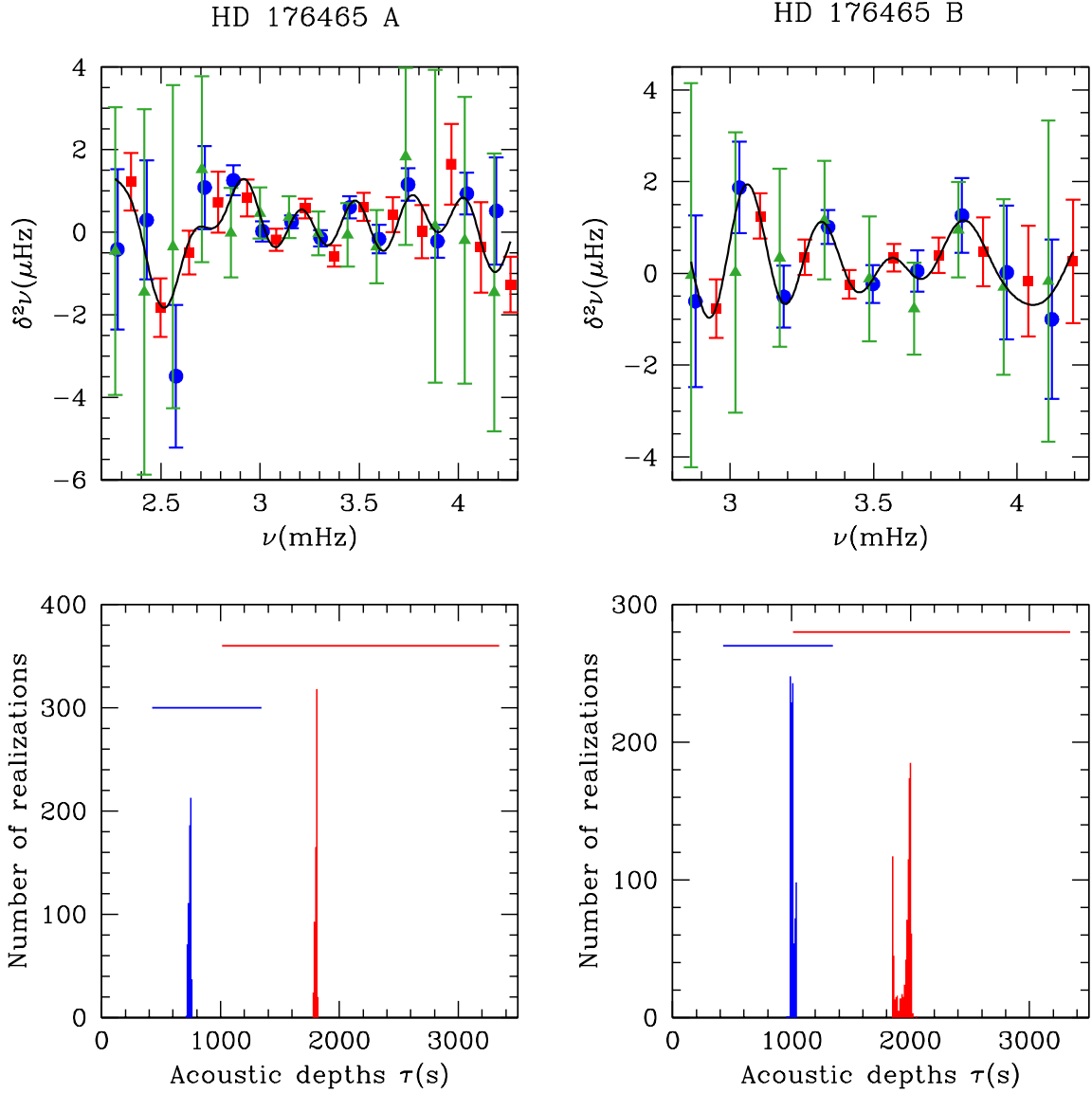}
\caption{Upper panels: fits to the second differences of the observed
frequencies for HD 176465 A (left) and B (right). The blue filled circles with
error bars are the second differences for $l$ = 0 modes, the red filled squares
are for $l$ = 1 modes and the green filled triangles are for $l$ = 2 modes. The
solid lines show the fit to the second differences using the form given by
Eq.(~\ref{eq:fsed}). Lower panels: the histograms of the fitted values of $\tau_{\rm He}$
(blue) and $\tau_{\rm CZ}$ (red) for different realizations. The horizontal
bars at the top show the range of initial guesses for the two parameters used
for fitting.}  \label{fig:fito}
\end{figure}

Using the technique described in \S~\ref{sec:technique}, we fit the signature
of acoustic glitches in the frequencies for both components of HD 176465 to
estimate the helium abundance, the acoustic depths of the He{\sc ii} ionization
zone $\tau_{\rm He}$ and the base of the convection envelope $\tau_{\rm CZ}$.
The upper panels of Figure~\ref{fig:fito} show the fitting of the observations.
The error bars of the second differences for $l=2$ mode are larger than that for
$l=0$ and $l=1$ modes. This is most likely to be a result of the larger
uncertainties of the $l=2$ mode frequencies.  The bottom panels show the
histograms of the fitted values of $\tau_{\rm He}$ (blue) and $\tau_{\rm CZ}$
(red). For HD 176465 A, $\tau_{\rm He}$ and $\tau_{\rm CZ}$ are $742_{-15}^{+10}$ s
and $1806_{-12}^{+6}$ s respectively. For HD 176465 B, $\tau_{\rm He}$ and
$\tau_{\rm CZ}$ are $1006_{-15}^{+28}$ s and $1979_{-125}^{+21}$ s
respectively. The acoustic depth of the glitches for HD 176465 A are shallower
than those of component B.  This is expected because the effective temperature
$T_{\rm eff}$ of component B is smaller than that of component A, and cooler
stars have deeper convection zones, as well as deeper helium ionization layers
(see also Verma et al. 2014b).

Besides the acoustic depth are estimated by fitting the second differences
of the observations, we also calculate $\tau_{\rm He}$ and $\tau_{\rm CZ}$ from
the known sound speed profile of the best-fit models (M8A and M8B in
Table~\ref{tab:bestfit}) by $\tau$ = $\int_{r}^{R}dr/c$. $r$ is the radial
distance of the He{\sc ii} ionization zone or the base of the convection zone.
Based on the sound speed profile, we get $\tau_{\rm He}$ = 702 s and $\tau_{\rm CZ}$
= 1990 s for HD 176465 A, $\tau_{\rm He}$ = 661 s and $\tau_{\rm CZ}$ = 1899 s for
HD 176465 B. From comparison of the results with two different methods, we can see
that the two estimations of $\tau_{\rm CZ}$ match well for both components of HD176465.
$\tau_{\rm He}$ are consistent with each other within 2.5$\sigma$ for HD176465A.
But for a lower mass star HD176465 B, there is a significant difference between the two
estimations of $\tau_{\rm He}$. This difference maybe result from the uncertainty
in the definition of the stellar surface (Houdek \& Gough 2007) and the ambiguity
in the position of the He{\sc ii} ionization zone (Verma et al. 2014b).
As can be see from the above comparison, the robustness of fit primarily depends on
the mass of star. This issue has also been discussed by Verma et al. (2017).
Because HD176465 A and B are the sub-solar mass stars that are not
too low in mass, the fit to the CZ signatures are robust.
For low-mass stars, the depression in $\Gamma_{1}$ profile in He{\sc ii} ionization
zone is shallow, which result in the small amplitude of the helium signature.
Especially for the sub-solar mass stars, it is difficult to fit the helium signature
because of the small amplitude, unless sufficiently low radial order modes are observed.
Consequently, the higher mass and more sufficient low radial order
frequencies ($n$ = 13 $\sim$ 28 of $l$ = 0, 1 and $n$ = 12 $\sim$ 27 of $l$ = 2)
for HD176465A than those ($n$ = 16 $\sim$ 26 of $l$ = 0, 1 and
$n$ = 15 $\sim$ 25 of $l$ = 2) for HD176465 B, make the better quality of fit for
primary component, which also justify the importance of the low radial order modes
for improve the quality of the fitting.

\begin{figure}
\epsscale{1.0} \plotone{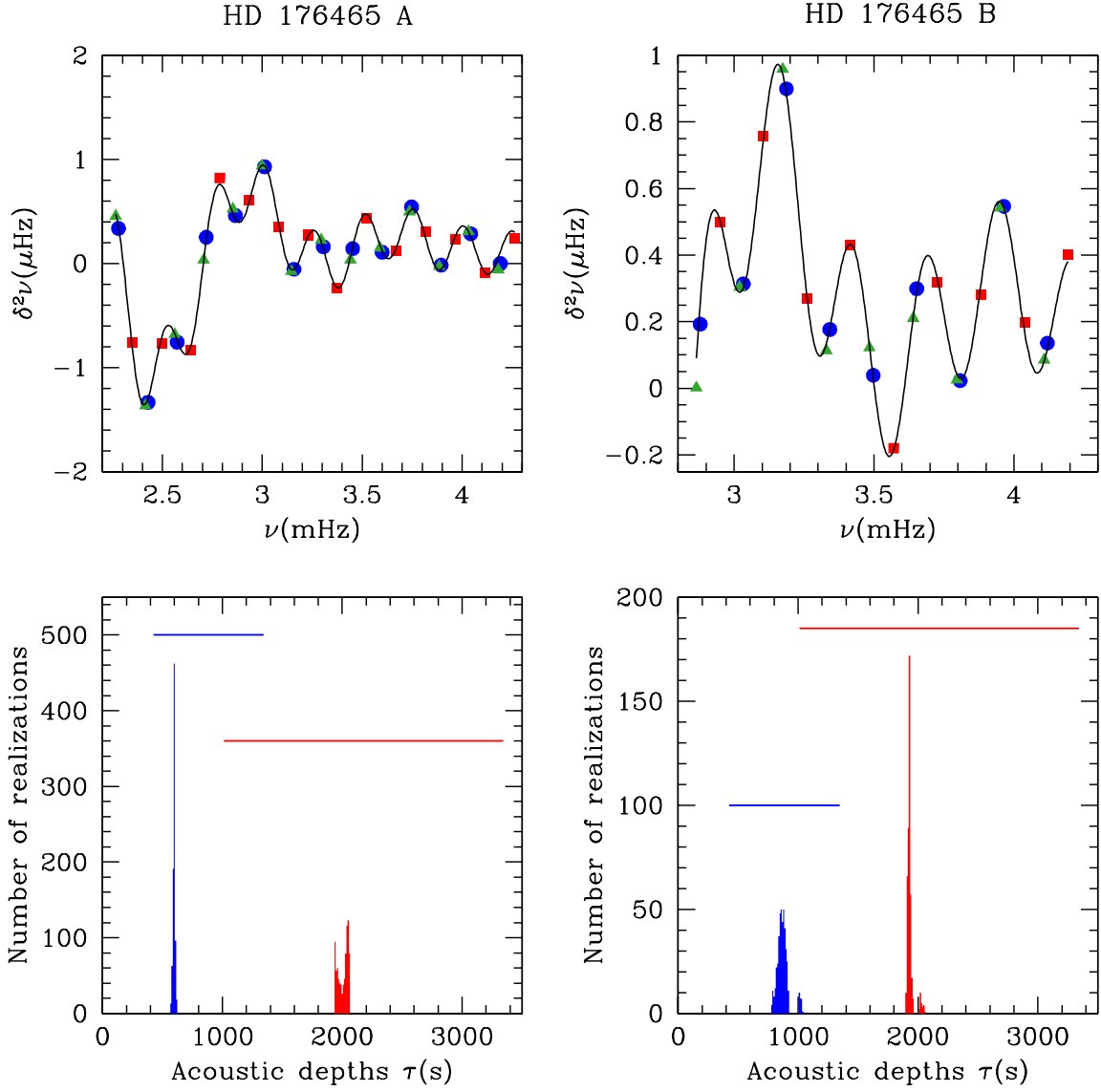}
\caption{Fit to the second differences for the best-fit models M8A and M8B for HD 176465 A
(left) and HD 176465 B (right). The symbols are same as in Figure~\ref{fig:fito}}
\label{fig:fitm}
\end{figure}

In Figure~\ref{fig:fitm} we show the results for our best-fit models (M8A and
M8B in Table~\ref{tab:bestfit}), and we fit the second differences
derived from the individual frequencies. The selected range of the theoretical
frequencies for fit are the same as the observed frequencies (i.e. $n$ = 13
$\sim$ 28 of $l$ = 0, 1 and $n$ = 12 $\sim$ 27 of $l$ = 2 for HD176465A; $n$ =
16 $\sim$ 26 of $l$ = 0, 1 and $n$ = 15 $\sim$ 25 of $l$ = 2 for HD176465B).
Comparing the fit for components A with B, we find that the
fit to the CZ signature for lower-mass component B is more robust than that
for component A, but that is not the case for the fit to the helium signal.
The signature from the He{\sc ii} ionization layers is weaker for component
B which has a lower amplitude of the He signal.  The distribution of the
fitted parameter $\tau_{\rm He}$ have multiple peaks. We select those
realizations for which the fitted acoustic depth $\tau_{\rm He}$ falls in the
dominant peak to calculate the median and the error estimations. As mentioned
above, the smaller number of low radial-order modes, and the larger uncertainties
in the frequencies for component B make it difficult to fit the He signature.
Another complicating factor is that as the mass of a star decreases, the dip
in $\Gamma_1$ for a given helium abundance decreases too (see also Verma et al. 2014b).

In order to determine the helium abundance of HD 176465, we select all models
of component A with $\chi_{\nu,{\rm A}}^{2} < 50$;  we refer to this set of
models as Set~A. For component B we select models with $\chi_{\nu, {\rm B}}^{2}
< 10$ (Set~B). We fit the various acoustic glitch parameters for Set~A
using the same range of frequencies and errors as the observations for
component A. We do a similar analysis for models in Set~B, but using the
observed mode-set of component B. We compare the observed
amplitudes of the He{\sc ii} ionization signature to those for the models of
Set~A and Set~B; the models have known helium abundances and hence, can be
used to calibrate the amplitude of the He{\sc ii} ionization signature.
Furthermore, from the models in Set~A and B, we select
the models with average amplitude of the He{\sc ii} ionization zone signature
consistent with that derived from the observed frequencies within 1$\sigma$
uncertainty, and define these models as Subset~A and Subset~B to estimate
the helium abundance of HD176465 A and B respectively.
The $\chi^{2}_{\nu}$ limits were adopted to yield a reasonable number
of models to statistically analyse the helium abundance of stars; since component
B has larger errors in frequencies, there are more models with lower
$\chi^{2}_{\nu}$. Additionally, the large separations $\Delta\nu$ are in the range
146.83$\sim$147.17$\mu Hz$ for models in Subset~A and 155.09$\sim$155.49$\mu Hz$ for
models in Subset~B, which are consistent with the observations ($\Delta\nu_{A}$ =
146.79 $\pm$ 0.12 $\mu Hz$, $\Delta\nu_{B}$ = 155.42 $\pm$ 0.13 $\mu Hz$) within 2$\sigma$.
Hence, the choice of the thresholds for $\chi^{2}_{\nu}$ also avoid the significant
deviation for the asteroseismic properties between models and observation.

\begin{figure}
\epsscale{1.0} \plotone{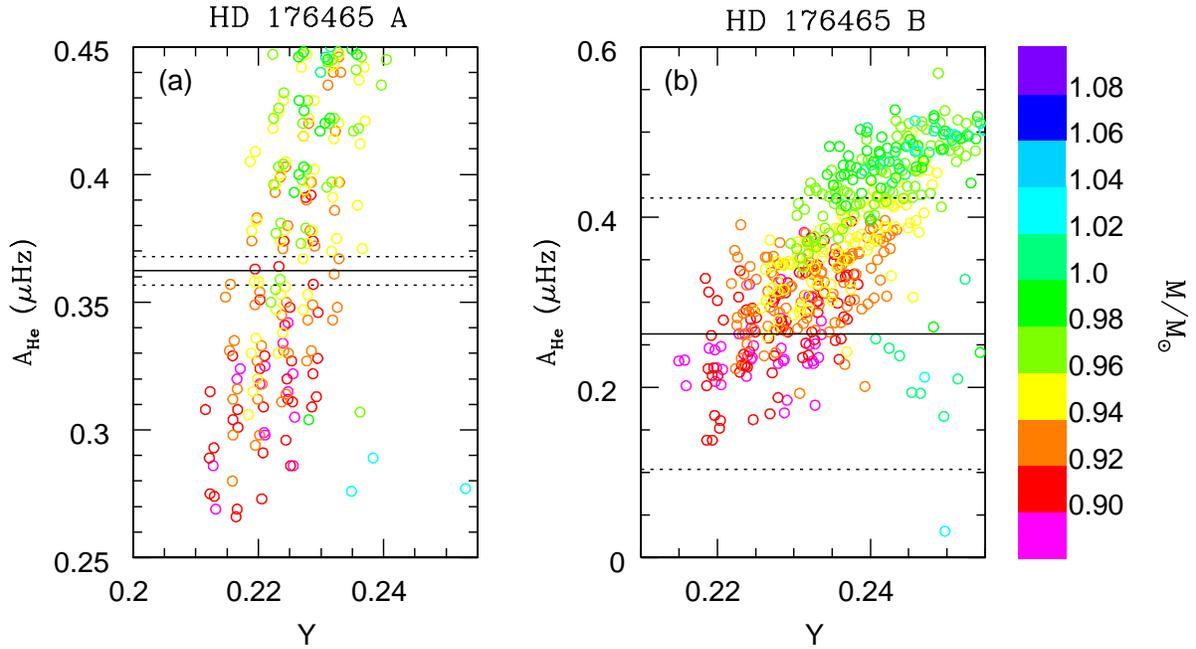}
\caption{The average amplitude of the oscillation signature due to the He{\sc ii}
ionization zone of models in Sets~A and B as a function of the current helium abundance in the
envelope. The open circles represent the results of the models color-coded by
mass. The horizontal solid line is the amplitude
from observation. The dotted lines are the 1$\sigma$ uncertainty of the
observed results. }  \label{fig:avy}
\end{figure}

Figure~\ref{fig:avy} shows the average amplitude of acoustic
glitch oscillation signature from the second helium ionization zone
for the observations and each selected model. In the figure, the
amplitude of the He{\sc ii} ionization zone signature for the observations
is represented by the horizontal solid line and 1$\sigma$ uncertainty
by the dotted lines. The amplitude of the He{\sc ii} ionization
zone signature for the observations is the median value from the
distribution of 1001 average fitted amplitudes for 1000 realizations
and actual observed frequencies. The upper 34\% and lower 34\% of
the distribution on either side of the median is the $\pm$ 1$\sigma$
uncertainty of the amplitude. The open circles with different
colors show the amplitude of the models with various masses. Since
the amplitude is frequency dependent, the amplitudes for the
observation and models are averaged over the same fitting frequency
interval. From the results of the models plotted in Figure~\ref{fig:avy},
we can see that, as expected, the amplitude of the He{\sc ii}
ionization zone signature increases with the amount of helium in
the envelope. Moreover, the stellar mass and the initial helium
abundance are anti-correlated (Lebreton \& Goupil 2014, Silva Aguirre et al.
2015). Hence, there is some scatter due to variations in other parameters,
for instance mass. Figure~\ref{fig:avy} show
that, for a given helium abundance, the amplitude
also increase with the mass of the star, a result of the fact that
the decrease in $\Gamma_1$ for a given helium abundance increases
with mass for our mass range. For component B, the amplitude of
observation is smaller than that for component A, and the uncertainty of
amplitude are much larger than component A by a factor of ten or more.
Figure~\ref{fig:fito} shows that the number of the observed frequencies for
component B is less than that for component A. Furthermore, the errors in
the second differences are very large. These factors make it difficult
to fit the  He{\sc ii} signature of component B. Verma et al.  (2014a)
have demonstrated that the precision of determination the helium abundance can
be improved significantly by adding more low order mode frequencies or
improving the precision of these frequencies. In order to understand
this improvement, we repeat the fit of second differences of observed
frequencies after removing successively the lowest three order modes
of degree $l$ = 0, 1, 2 for only HD176465 A. We find that the uncertainty in
the amplitude of the He{\sc ii} ionization zone signature increases
rapidly by a factor of ten or more. After removing the lowest order
modes for HD176465 A, the uncertainty in amplitude (0.174 $\mu Hz$)
is similar with that of HD176465 B (0.159 $\mu Hz$). In this case,
the lowest order modes ($n$ = 16 of $l$ = 0,1 and $n$ = 15 of $l$ = 2)
in fitting for HD176465 A are the same as those for HD176465 B.
The amplitude of He signal is larger using the low order mode
frequencies which are important for stabilizing the fitting of
the helium signature.

Finally, we obtain the helium
abundance for both components by calibrating the amplitude of models with
observations. From the models in Subset~A and B, as mentioned above,
with known helium abundance, we evaluate the mean
value of helium and the standard deviation of the models as the helium of
HD176465 A and B, $Y_{\rm A} = 0.224 \pm 0.006$ and $Y_{\rm B} = 0.233
\pm 0.008$. The current helium abundances for HD 176465 A and B are
consistent with each others within errors, the small difference could
be a result of differences in settling efficiency because of the
difference in masses of the two components.

\subsection{Ensemble Analysis}
\label{sec:ensemble}

W use the models in Sets A and B to study the variations of the acoustic
glitch signature in frequencies as a function of the different stellar parameters.  The
large frequency separations of the models, which are calculated from the
surface-term corrected frequencies, are in the range $146.7 \leq \Delta\nu \leq
147.2\;\mu$Hz.  Models in Set~B have large separations in the range $155.1 \leq
\Delta\nu \leq 155.5\;\mu$Hz

\begin{figure}
\epsscale{1.0} \plotone{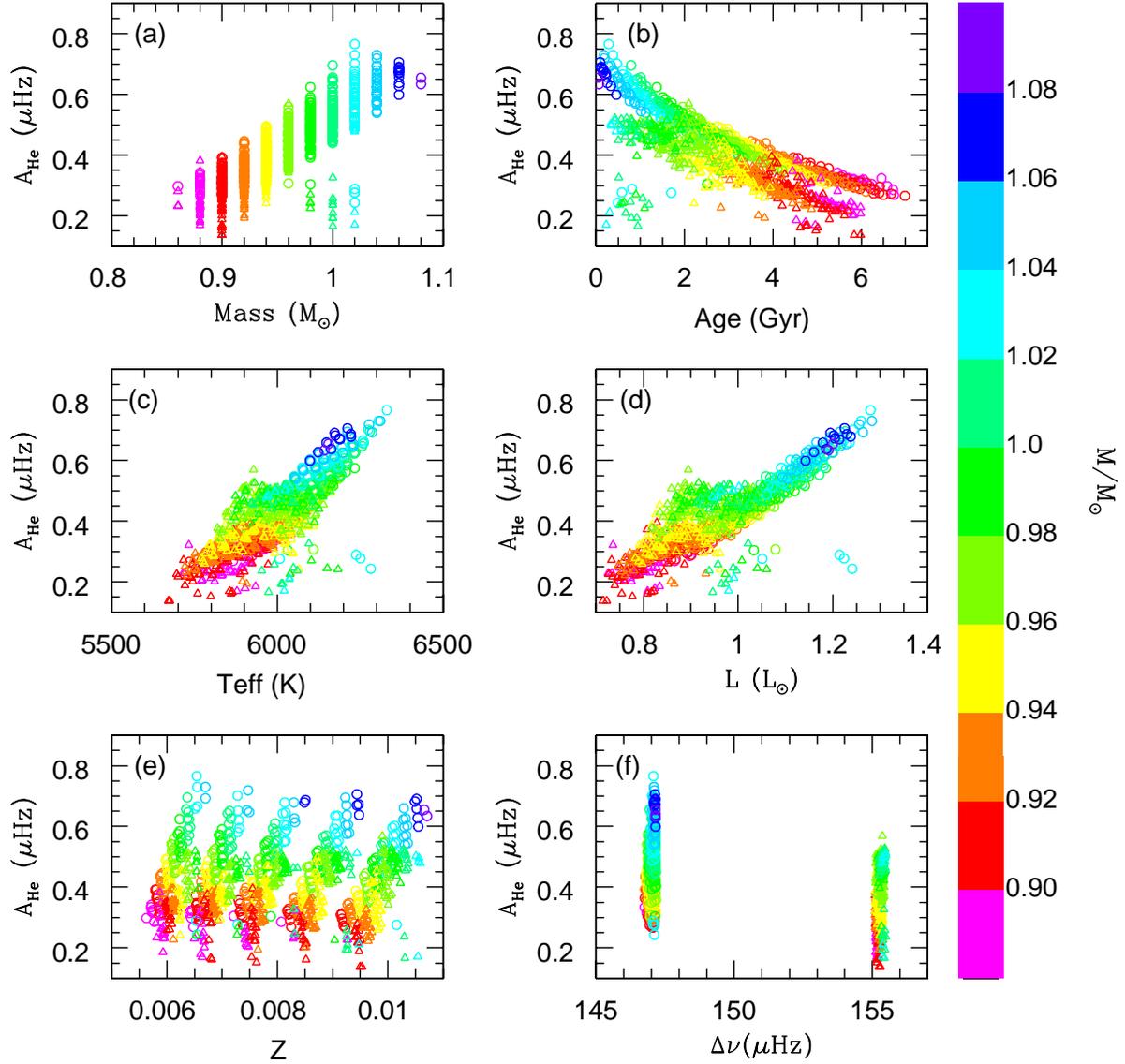}
\caption{The variation of the average amplitude of He{\sc ii} signature with
mass, age, effective temperature, luminosity, metallicity and the large
separation. In each panel the open circle represent models in
Set~A and open triangles are models in Set~B.
The different colors represent the different masses of the models.}
\label{fig:avhe}
\end{figure}

\begin{figure}
\epsscale{1.0} \plotone{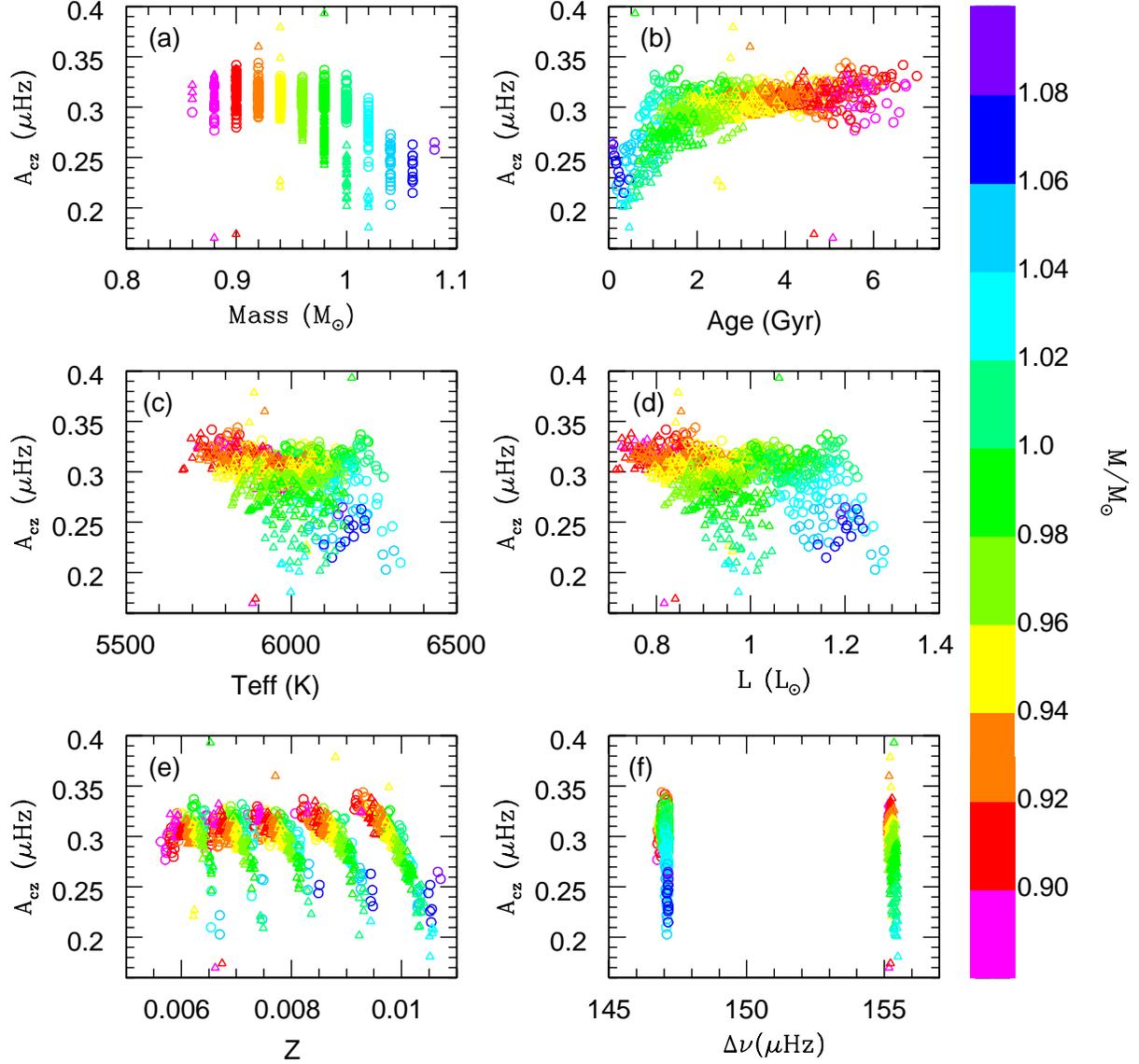}
\caption{The same as Figure~\ref{fig:avhe}, but
for results of the average amplitude of the base of convection zone.
}  \label{fig:avc}
\end{figure}

Figure~\ref{fig:avhe} shows the dependence of the average amplitude of He{\sc ii}
ionization zone signature $A_{\rm He}$ in the various stellar parameters for
Set~A (open circles) and Set~B (open triangles). $A_{\rm He}$ increase as a function
of the mass as was also seen in Figure~\ref{fig:avy}. The change of $A_{\rm He}$
with effective temperature $T_{\rm eff}$ and luminosity $L$ is a reflection of the
change in mass. The amplitude $A_{\rm He}$ decreases with age not only due to the
lower mass with older star for a given $\Delta\nu$ but also due to the reduced
helium abundance in the envelope by the helium diffusion as the star evolves.
The effect of diffusion is clearly shown in panel (e). For a given initial
metallicity, the lower mass star with older age has lower metallicity. For a
given mass, $A_{\rm He}$ decrease with metallicity slightly. Figure~\ref{fig:avc}
shows the variation of the average amplitude of the signature from the
 base of the convective
envelope, $A_{\rm CZ}$, as a function of the different parameters.
$A_{\rm CZ}$ decrease with increase in
mass, effective temperature, luminosity, metallicity and the large frequency
separation, while it increase with age. The variation of $A_{\rm CZ}$ is quite
modest for sub-solar mass models, but for masses larger than 1.0$M_{\odot}$,
$A_{\rm CZ}$ changes more rapidly.

\begin{figure}
\epsscale{1.0} \plotone{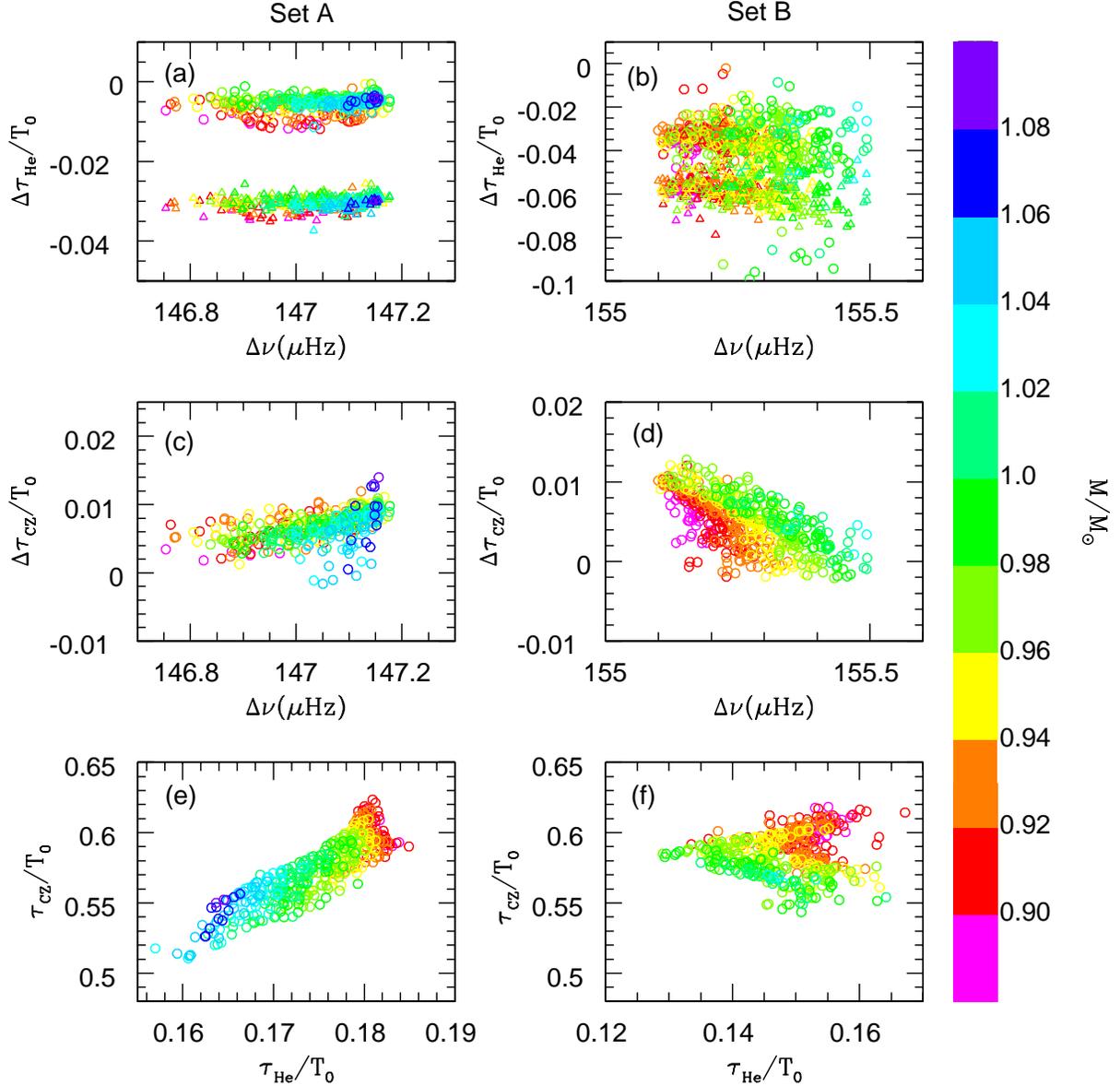} \caption{
Panels (a) (b) show the differences of the acoustic depth of He{\sc ii}
ionization zone between the fitted results and obtained using the sound
profile for Sets~A and B respectively.
The selected models are same as Figure~\ref{fig:avhe}. The open
circles represent the scaled differences $\Delta\tau_{\rm He}/T_{0} = (\tau^{\rm
fit}_{\rm He} - \tau^{c}_{He,\Gamma_{1,peak}})/T_{0}$. The open triangles
represent the scale differences $\Delta\tau_{He}/T_{0} = (\tau^{fit}_{He} -
\tau^{c}_{He,\Gamma_{1,dip}})/T_{0}$. Panels (c) and (d) show the scaled differences
of the acoustic depth of the base of convection zone between the fitted results
and obtained using the sound profile. Panels (e) and (f) show the scaled acoustic
depth of the base of convection zone as a function of that of He{\sc ii}
ionization zone.}  \label{fig:dtau}
\end{figure}

The study of acoustic-glitch signatures in oscillations allow us not only to estimate the
helium abundance but also to detect the location of the glitches. Figure~\ref{fig:dtau}
shows the results for Set~A (left column) and Set~B (right column). In
Figure~\ref{fig:dtau}, the top panels (a) and (b) show the scaled differences of
the acoustic depth $\Delta\tau_{\rm He}/T_{0}$ between the result obtained from the
fitting analysis $\tau^{\rm fit}_{He}$ and the result from calculating the sound
speed profile of the model, i.e.,
\begin{equation}
\tau^{c}_{\rm He}=\int^{R}_{r_{\rm He}} \frac{dr}{c},
\label{eq:tauhe}
\end{equation}
where $R$ is the radius of the star, $r_{He}$ is the radial distance where the
He{\sc ii} ionization zone is located, and $c$ is the sound speed. In panels (a) (b),
open circles represent the scale differences $\Delta\tau_{\rm He}/T_{0} =
(\tau^{\rm fit}_{\rm He} - \tau^{c}_{{\rm He},\Gamma_{1,peak}})/T_{0}$, where
$\tau^{\rm fit}_{\rm He}$ is the value of the acoustic depth obtained by
fitting the second differences.
The open triangles
represent the scale differences $\Delta\tau_{\rm He}/T_{0} = (\tau^{\rm fit}_{\rm He} -
\tau^{c}_{{\rm He},\Gamma_{1,{\rm dip}}})/T_{0}$.
$\tau^{c}_{{\rm He},\Gamma_{1,{\rm peak}}}$ is the
acoustic depth corresponding to the He{\sc ii} ionization zone of peak in the
first adiabatic index $\Gamma_{1}$ profile between the He{\sc i} and He{\sc ii}
ionization.  $\tau^{c}_{{\rm He},\Gamma_{1,{\rm dip}}}$ is the acoustic depth
of the location of the dip in $\Gamma_{1}$. $T_{0}$ is the total acoustic
radius, which are calculated from the average large separation of radial
frequencies $T_{0} \approx (2 \Delta\nu_{0}) ^{-1}$. From panels (a) and (b),
we can see that the acoustic glitch of the helium ionization layer match with
the acoustic depth of the peak in $\Gamma_{1}$ profile, a feature seen earlier
by Broomhall et al. (2014) and Verma et al. (2014b). In Figure~\ref{fig:dtau},
panels (c) and (d) show the scaled
difference of the acoustic depth $\Delta\tau_{\rm CZ}/T_{0}$ between the
fitting analysis and the calculated result from the sound profile and the
differences almost less than 0.01. Comparing the results between the He
signature [panels (a) and (b)] and the CZ signature [panels (c) and (d)], we
find that the quality of fit has primary dependence on mass. For the sub-solar
and solar mass models, the fitting
analysis to the CZ signal was robust, while the fit to the He{\sc ii} ionization
signal was not good especially for the models with mass less than 0.9
$M_{\odot}$ due to the smaller amplitude of the peak in $\Gamma_{1}$ profile
between the He{\sc i} and He{\sc ii} ionization.
The lowermost panels (e) and (f) of Figure~\ref{fig:dtau} show the clear correlation
between the scale acoustic depth of the base of convection envelope and the second
helium ionization zone especially for Set~A, but not for models with mass
less than 0.9 $M_{\odot}$ again due to the difficulty of fitting for lower mass
star. The positive correction between $\tau_{\rm CZ}/T_{0}$ and $\tau_{\rm He}/T_{0}$
show that the cooler star with lower mass have the deeper convection envelope
and the second helium ionization zone.

\begin{figure}
\epsscale{1.0} \plotone{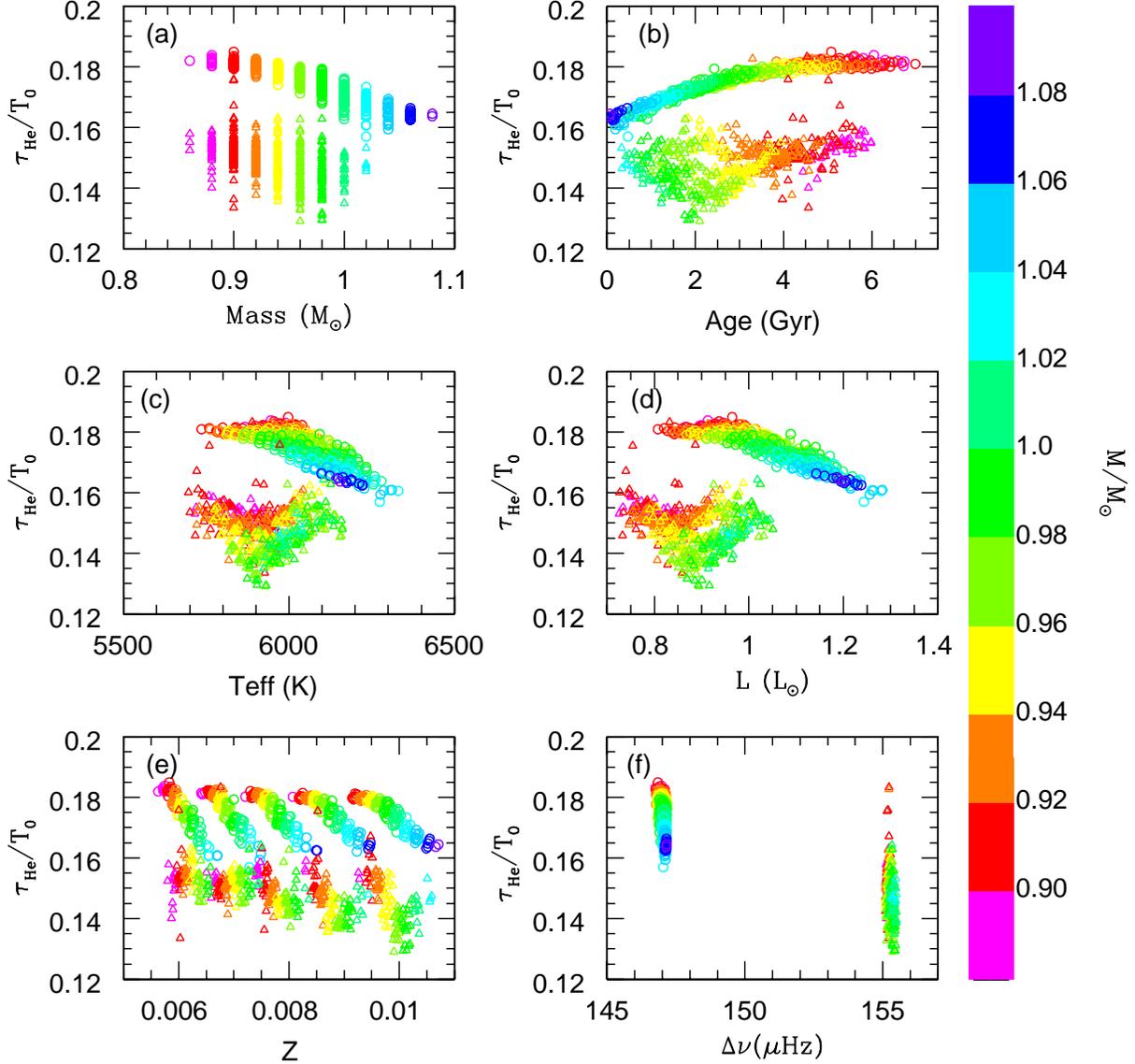}
\caption{The variation of the scaled acoustic depth of He{\sc ii} ionization
zone with mass, age, effective temperature, luminosity, metallicity and the
large separation respectively. The selected models are same as
Figure~\ref{fig:avhe}. In each panel of figure, the open circle represent the
Set~A which calibrated by the observations of component A. The open triangle
represent the Set~B which calibrated by the observations of component B.
Different colors show the different masses of stellar models.}
\label{fig:tauhe}
\end{figure}

Figure~\ref{fig:tauhe} shows the scaled acoustic depth of He{\sc ii} ionization zone
signature $\tau_{\rm He}/T_{0}$ from fitting analysis as a function of mass,
age, $T_{\rm eff}$, $L$, $Z$ and $\Delta\nu$. $\tau_{\rm He}/T_{0}$ decrease with the
increased mass, $T_{\rm eff}$, $L$, $Z$ and $\Delta\nu$, while it increase as the
star evolve. This is again show that the hotter star with larger mass has the
shallower helium ionization zone. Comparing the results for Set~A (open
circles) and Set~B (open triangles), the larger dispersion of the result for
Set~B show that more sufficient and precise low radial mode observed
frequencies of component B are needed to improve the quality of the fits to Set~B.

\begin{figure}
\epsscale{1.0} \plotone{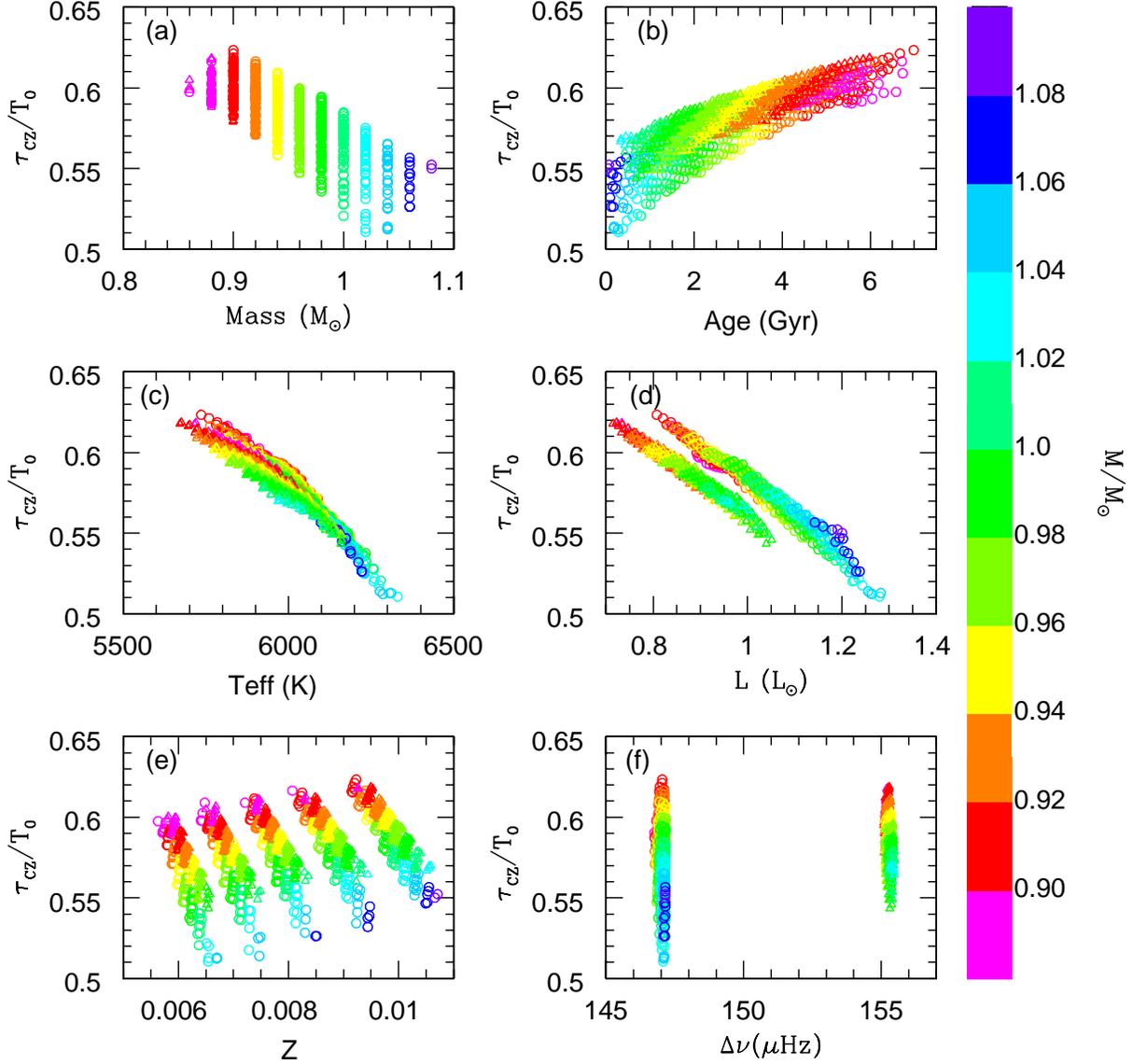}
\caption{The variation of the scaled acoustic depth of the base of convection
zone with mass, age, effective temperature, luminosity, metallicity and the
large separation respectively. The selected models and symbols are same as
Figure~\ref{fig:avhe}.}  \label{fig:tauc}
\end{figure}

Figure~\ref{fig:tauc} shows that there are clear dependence of $\tau_{\rm CZ}/T_{0}$
on various stellar parameters for both Sets~A and B. Comparing the results in
Figure~\ref{fig:tauc} with Figure~\ref{fig:tauhe}, we find that the fit to the
base of the convective envelope is more robust than the fit to the He{\sc ii} ionization signal for
sub-solar mass stars. In each panel of Figure~\ref{fig:tauc}, the vertical
distribution of $\tau_{\rm CZ}/T_{0}$ show that the dependence of $\tau_{\rm CZ}/T_{0}$
on mass is significantly greater than on other parameters. For a given mass,
poorer metallicity models have lower opacity, resulting in a shallower convective
envelope.

\section{Conclusions}
\label{sec:conc}

We have used observed mode frequencies and spectroscopic parameters to derive
the stellar parameters for binary system HD 176465. The masses and radii of
both components are consistent with the earlier estimates of White et al.
(2017) using different codes. Although we fit the observations of HD~176465
A and B individually, we find that both models have the same initial abundances
($Y_{\mathrm{init},{\rm A}}=0253\pm 0.006$ and $Y_{\mathrm{init}, {\rm B}}=0.254\pm 0.008$;
$Z_{\mathrm{init}}=0.010\pm 0.001$ for both components).
 The current metallicity of the two components  are also very similar
([Fe/H]$_{\rm A} = -0.275 \pm
0.04$ and [Fe/H]$_{\rm B} = -0.285 \pm 0.04$);
note these are lower than the initial
metallicity because of the gravitational settling of element diffusion. Although the
ages of two stars of HD 176465 are not a priori constrained to be the same, the
age of both components ($t_{\rm A} = 3.087 \pm 0.580$ Gyr and $t_{\rm B} = 3.569 \pm
0.912$ Gyr) are the same within errors. This show that both
components of  HD~176465 are the young sub-solar mass stars, with masses
of $0.975\pm 0.040$ M$_\odot$ and $0.932\pm 0.022$ M$_\odot$.

Additionally, we used the glitch signal in the frequencies caused by the
depression in $\Gamma_{1}$ at the second helium ionization zone to
determine the current envelope helium abundance of the stars.
We used the technique of fitting the second differences in
frequencies to the function proposed by Basu et al. (2004).
However, the exact form of the amplitudes of glitch signatures or the different techniques to remove the
smooth term in the frequencies does not affect the results significantly (Basu
et al. 2004; Verma et al. 2014a). We estimate the current helium abundance
($Y_{\rm A} = 0.224 \pm 0.006$ and $Y_{\rm B} = 0.233 \pm 0.008$) for both components,
which are lower than the initial helium abundance because of the effect of
element diffusion. The small difference of helium depletion is consistent
with what is expected from the small difference in mass between the two
components of the binary.

We selected two sets of models whose asteroseismic properties are
very close to those of the two components of HD 176465
to study the properties of the acoustic glitch signatures arising
from the second helium ionization layer and the base of the convective
envelope. The amplitude of He{\sc ii} signature $A_{\rm He}$ is predominantly a function
of the helium abundance. $A_{\rm He}$ also increase with mass, effective
temperature, luminosity and decrease as the star evolves. The variation of the
amplitude of the base of convection zone ($A_{\rm CZ}$) is quite modest for
sub-solar mass models. Furthermore, we determined the location of the He and CZ
signatures and confirmed that acoustic depth of the helium ionization
layer matches the
acoustic depth of the peak in $\Gamma_{1}$ profile between the He{\sc i} and He{\sc ii}
ionization zones. The fit to the CZ signature are robust for low-mass star with
mass less than 1.0 $M_{\odot}$. The acoustic depth of convection zone
$\tau_{\rm CZ}$ is sensitive to the metallicity because of the poorer metallicity
have lower opacity, hence shallower convective envelopes. The dependence of
$\tau_{\rm CZ}$ on the metallicity can be used to constrain the metallicity. The
fit of the He signature are weak for the models with mass less than 0.9
$M_{\odot}$ because of the smaller amplitude of the peak in $\Gamma_{1}$. The
acoustic depth between $\tau_{\rm CZ}$ and $\tau_{\rm He}$ present a clear positive
correlation, which show that the deeper convection envelope and the deeper
second helium ionization zones for the cooler low mass star. The robust fit of
CZ signature for lower mass star and the positive correlation between
$\tau_{\rm CZ}$ and $\tau_{\rm He}$ can help us in estimating  $\tau_{\rm He}$ for stars
with masses less than 0.9 $M_{\odot}$. Besides the effect of mass, the more
sufficient and precise low radial mode observed frequencies can help to improve
the quality of the fit.

\acknowledgments
We thank the referee for constructive comments that led to significant
improvements of this paper. N.G. and Y.K.T. acknowledge Grant No. 11673005
from the National Natural Science Foundation of China. S.B. acknowledges
NSF grant AST-1514676 and NASA grant NNX16AI09G.

\begin{sidewaystable}
\scriptsize
\centering
\caption{\scriptsize{Properties of the better candidate models for HD 176465 A and B}}
\label{tab:bestfit}
\begin{tabular}{c c c c c c c c c c c c c c c c c}
\hline\hline
star&$\alpha_{\mathrm{MLT}}$&M&Age&Y$_{\mathrm{int}}$&Z$_{\mathrm{int}}$&[Fe/H]&R&T$_{eff}$&$\Delta\nu$&$\nu_{max}$
&$\chi^{2}_{T_{eff}}$&$\chi^{2}_{[Fe/H]}$&$\chi^{2}_{\nu}$&$\chi^{2}$&$a_{\mathrm{cubic}}$&$a_{\mathrm{inv}}$\\
HD176465&&(M$_{\odot}$)&(Gyr)&&&&(R$_{\odot})$&&($\mu Hz$)&($\mu Hz$)&&&&&&\\
\hline
\multicolumn{17}{c}{(Z/X)$_{\odot}$=0.019(AGSS09)}\\
\hline
M1A&1.6&0.94&3.245&0.254&0.0084&-0.3612&0.922&5860&147.08&3389& 0.114&1.041&6.704& 7.859&3.245&-11.203\\
M1B&1.6&0.90&3.108&0.258&0.0067&-0.4594&0.876&5871&155.26&3597& 2.127&7.054&0.948&10.129&3.108&-11.167\\
\hline
M2A&1.7&0.96&2.761&0.254&0.0084&-0.3556&0.929&5957&147.07&3388& 1.984&0.858&6.272& 9.114&2.761&-11.205\\
M2B&1.7&0.92&2.798&0.254&0.0067&-0.4548&0.883&5945&155.27&3598& 5.194&6.659&0.989&12.842&2.798& -9.176\\
\hline
M3A&1.826&0.98&2.596&0.250&0.0085&-0.3529&0.935&6037&147.05&3386& 5.311&0.778&6.499&12.588&2.596& -9.451\\
M3B&1.826&0.94&2.660&0.258&0.0084&-0.3486&0.889&5954&155.28&3619& 5.655&0.656&1.040& 7.351&2.660&-10.323\\
\hline
M4A&1.9&0.96&3.170&0.258&0.0084&-0.3618&0.929&6064&147.04&3358& 6.745&1.060&6.531&14.336&3.170&-10.552\\
M4B&1.9&0.94&2.760&0.250&0.0068&-0.4538&0.889&6053&155.25&3594&12.083&6.567&0.965&19.615&2.760& -9.678\\
\hline
M5A&2.0&0.98&2.869&0.254&0.0084&-0.3574&0.935&6127&147.02&3362&10.913&0.915&6.165&17.993&2.869& -9.138\\
M5B&2.0&0.94&2.827&0.254&0.0067&-0.4556&0.888&6111&155.23&3578&17.016&6.729&1.068&24.813&2.827& -9.336\\
\hline
\multicolumn{17}{c}{(Z/X)$_{\odot}$=0.023(GS98)}\\
\hline
M6A&1.6&0.96&3.220&0.250&0.0107&-0.2515&0.930&5756&147.14&3434& 0.671&0.652&5.797& 7.120&3.220&-10.011\\
M6B&1.6&0.90&3.114&0.258&0.0067&-0.4563&0.876&5867&155.25&3599& 2.006&6.783&0.969& 9.758&3.114&-11.869\\
\hline
M7A&1.7&0.96&3.181&0.258&0.0106&-0.2531&0.930&5854&147.11&3410& 0.069&0.610&5.439& 6.118&3.181&-10.677\\
M7B&1.7&0.92&2.801&0.254&0.0068&-0.4517&0.882&5941&155.27&3601& 5.009&6.394&0.953&12.356&2.801&-10.076\\
\hline
M8A&1.826&0.98&3.227&0.250&0.0107&-0.2521&0.936&5913&147.12&3415& 0.853&0.638&5.298& 6.789&3.227&-10.578\\
M8B&1.826&0.96&2.802&0.250&0.0107&-0.2416&0.896&5827&155.34&3676& 0.935&0.947&0.994& 2.876&2.802&-10.251\\
\hline
M9A&1.9&0.98&3.139&0.250&0.0096&-0.3039&0.936&6007&147.07&3389& 3.893&0.004&5.589& 9.486&3.139& -9.083\\
M9B&1.9&0.94&2.766&0.250&0.0068&-0.4507&0.888&6049&155.25&3598&11.816&6.307&1.023&19.146&2.766&-10.593\\
\hline
M10A&2.0&1.00&2.579&0.258&0.0106&-0.2468&0.943&6079&147.07&3390& 7.704&0.787&5.527&14.018&2.579& -8.129\\
M10B&2.0&0.96&2.402&0.258&0.0085&-0.3428&0.896&6068&155.28&3607&13.321&0.508&1.081&14.910&2.402& -8.206\\
\hline\hline
\end{tabular}
\end{sidewaystable}

\begin{table}[h]
\scriptsize
\centering
\caption{\scriptsize{Comparison of our results (YREC1 and YREC2) with those obtained using other codes
(AMP, ASTFIT, BASTA and MESA) by white et al. (2017) for HD 176465 A.}}
\label{tab:compareA}
\begin{tabular}{c c c c c c c c}
\hline\hline
Property&AMP&ASTFIT&BASTA&MESA1&MESA2&YREC1&YREC2\\
\hline
M/M$_{\odot}$&0.930$\pm$0.04&0.952$\pm$0.015&0.960$^{+0.010}_{-0.011}$&0.95$\pm$0.03&0.99$\pm$0.02&0.975$\pm$0.016&0.944$\pm$0.018\\
Y$_{\mathrm{init}}$&0.258$\pm$0.024&0.262$\pm$0.003&0.265$\pm$0.002&0.25$\pm$0.02&0.23$\pm$0.02&0.253$\pm$0.006&0.256$\pm$0.005\\
Z$_{\mathrm{init}}$&0.0085$\pm$0.0010&0.0103$\pm$0.0010&0.011$\pm$0.004&0.0102$\pm$0.0009&0.0094$\pm$0.0009&0.0102$\pm$0.0008&0.0083$\pm$0.0003\\
Age(Gyr)&3.0$\pm$0.4&3.2$\pm$0.5&2.8$\pm$0.3&3.2$\pm$0.2&3.01$\pm$0.12&3.087$\pm$0.580&3.146$\pm$0.535\\
R/R$_{\odot}$&0.918$\pm$0.015&0.927$\pm$0.005&0.928$^{+0.006}_{-0.003}$&0.926$\pm$0.011&0.939$\pm$0.006&0.934$\pm$0.005&0.923$\pm$0.006\\
$\alpha_{\mathrm{MLT}}$&1.90$\pm$0.18&1.80&1.791&1.57$\pm$0.11&1.79&1.826&1.81$\pm$0.14\\
(Z/X)$_{\odot}$&0.023(GS98)&0.0245(GN93)&0.023(GS98)&0.023(GS98)&0.023(GS98)&0.023(GS98)&0.018(AGSS09)\\
\hline\hline
\end{tabular}
\end{table}

\begin{table}[h]
\scriptsize
\centering
\caption{\scriptsize{Comparison of our results (YREC1 and YREC2) with those obtained using other codes
(AMP, ASTFIT, BASTA and MESA) by white et al. (2017) for HD 176465 B.}}
\label{tab:compareB}
\begin{tabular}{c c c c c c c c}
\hline\hline
Property&AMP&ASTFIT&BASTA&MESA1&MESA2&YREC1&YREC2\\
\hline
M/M$_{\odot}$&0.930$\pm$0.02&0.92$\pm$0.02&0.929$^{+0.010}_{-0.011}$&1.02$\pm$0.07&0.97$\pm$0.04&0.932$\pm$0.022&0.914$\pm$0.021\\
Y$_{init}$&0.246$\pm$0.013&0.262$\pm$0.003&0.265$\pm$0.002&0.21$\pm$0.04&0.24$\pm$0.04&0.255$\pm$0.008&0.257$\pm$0.006\\
Z$_{init}$&0.0085$\pm$0.0007&0.0096$\pm$0.0011&0.011$\pm$0.004&0.0124$\pm$0.0015&0.0122$\pm$0.0011&0.0099$\pm$0.0008&0.0083$\pm$0.0004\\
Age(Gyr)&2.9$\pm$0.5&3.4$\pm$0.9&3.2$\pm$0.4&2.9$\pm$0.4&3.18$\pm$0.31&3.569$\pm$0.912&3.226$\pm$0.944\\
R/R$_{\odot}$&0.885$\pm$0.006&0.883$\pm$0.007&0.886$^{+0.003}_{-0.006}$&0.919$\pm$0.021&0.899$\pm$0.013&0.887$\pm$0.007&0.883$\pm$0.007\\
$\alpha_{\mathrm{MLT}}$&1.94$\pm$0.12&1.80&1.791&2.05$\pm$0.28&1.79&1.826&1.81$\pm$0.14\\
(Z/X)$_{\odot}$&0.023(GS98)&0.0245(GN93)&0.023(GS98)&0.023(GS98)&0.023(GS98)&0.023(GS98)&0.018(AGSS09)\\
\hline\hline
\end{tabular}
\end{table}

\newpage
\clearpage
\newpage
\clearpage

\newpage
\clearpage
\newpage
\clearpage

\newpage
\clearpage

\newpage
\clearpage

\newpage
\clearpage

\newpage
\clearpage

\newpage
\clearpage

\newpage
\clearpage

\newpage
\clearpage

\newpage
\clearpage
\end{document}